\documentclass[conference,a4paper]{IEEEtran}
\usepackage{xcolor}
\ifCLASSINFOpdf
 \usepackage[pdftex]{graphicx}
\else
\usepackage[dvips]{graphicx}
\fi
\ifCLASSOPTIONcompsoc
 \usepackage[caption=false,font=normalsize,labelfont=sf,textfont=sf]{subfig}
\else
 \usepackage[caption=false,font=footnotesize]{subfig}
\fi
\usepackage{siunitx}
\usepackage{url}
\usepackage{amsthm}
\usepackage{mathtools}
\usepackage{array}
\usepackage{cite}
\usepackage{tabularx}
\usepackage{amsmath,amssymb,amsfonts}
\usepackage{algorithmic}
\usepackage{graphicx}
\usepackage{textcomp}
\usepackage{bm}
\usepackage{siunitx}
\usepackage{dirtytalk}
 \usepackage{multicol}
\usepackage{breqn}
\begin{document}

\title{An Ultra-Wideband Microstrip MIMO Antenna with EBG Loading for WLAN and Sub-6G Applications}
\author{\IEEEauthorblockN{
Adel Omrani 
}                                 
\IEEEauthorblockA{IHM, KIT, Karlsruhe, Germany}
Email:  adel.hamzekalaei@kit.edu
}
\maketitle
\begin{abstract}
This manuscript presents an ultra-wideband Microstrip multiple-input–multiple-output (MIMO) antenna covering the $2.4 \sim 6.5$ GHz wireless local networks (WLAN) and Sub-6G bands. The applied MIMO antenna is composed of two symmetrical microstrip antenna elements. Each microstrip is designed based on the stepped impedance resonator (SIR) technique and closely positioned with about $0.25\lambda_0$ of the lower band (\SI{2.5}{GHz}). The proposed MIMO antenna is incorporated with the EBG structure to enhance the radiation characteristic and isolation between two radiating components. The return loss (${S_{11}}<–10$) dB covers $2.4 \sim 6.5$ GHz, and the proposed configuration obtains about \SI{20}{dB} isolation within the $2.4\sim 6.5$ GHz Sub-6G and WLAN bands, which shows an improvement compared to the initial design of the MIMO antenna. The envelope correlation coefficient (ECC) is under $0.02$ over the frequency band. The simulation results offer that the proposed microstrip MIMO antenna system with the different levels of the EBG structure is suitable for WLAN and Sub-6G applications. 
 \end{abstract}

\vskip0.5\baselineskip

\section{Introduction}
Multiple-input-multiple-output (MIMO) radio frequency (RF) architectures that incorporate IEEE standards 802.11a and 802.11b, can offer high data rates and high reliability by delivering higher receive gains and network capacities. It can address higher-speed internet services for handheld computers and smartphone users, high level of throughput, and good directivity for the local wireless networks (WLAN). Hence, many communication systems, such as 4G/5G technologies and WiFi systems,  have included the MIMO technology in the system specifications.\newline
\indent A crucial challenge in designing the MIMO WLAN system is fabricating compact, low-cost, relatively simple multiband, and broadband antennas whilst maintaining: $i)$ high isolation or interference between radiation elements, and  $ii)$ low envelope correlation. In recent years, various approaches, such as space/polarization/pattern diversity \cite{spacediv, Patterndiv, polardiv}, electromagnetic bandgap (EBG) structure \cite{EBG1}, T-slots plus a meander-line resonance \cite{Fconverted}, passive resonator, T-shape or L-shape ground branches \cite{Lshape}, and parasitic elements, to name a few, have been reported for reducing the mutual couplings in the design of MIMO antennas.\newline
One solution to reduce the mutual coupling is using dual-band decoupling networks like neutralization lines \cite{Neutral}, T-stub circuits \cite{Tstub}, and lumped components \cite{Lumped} by placing them between the radiating elements, which increases the complexity of the antenna. Ground plane structure (DGS) is another newly introduced method for coupling reduction between the antenna's radiating elements \cite{DGS1, DGS2}. In this method, the bottom layer of the antenna contains the DGS etched out from the ground plane. High isolation, however narrowband in the desired frequency band, is reported by employing this method. In \cite{Fconverted}, a meandering resonant branch and an inverted T-shaped slot as a parasitic element are used for the higher and lower bands to achieve high isolation properties. \newline
\indent In this letter, a single microstrip antenna is designed using the SIR technique. Next, two single antennas with bend transmission lines are united to construct the MIMO antenna. Later, a new design that makes use of a metamaterial-inspired EBG is proposed to improve the mutual coupling between the radiating elements, and the radiation characteristics. This MIMO antenna is operating in the $2.4 \sim 6.5$ GHz WLAN frequency range. The detail of the proposed antenna is discussed in the following.
\begin{figure}[!t]
\centering
\includegraphics[width=0.5\textwidth]{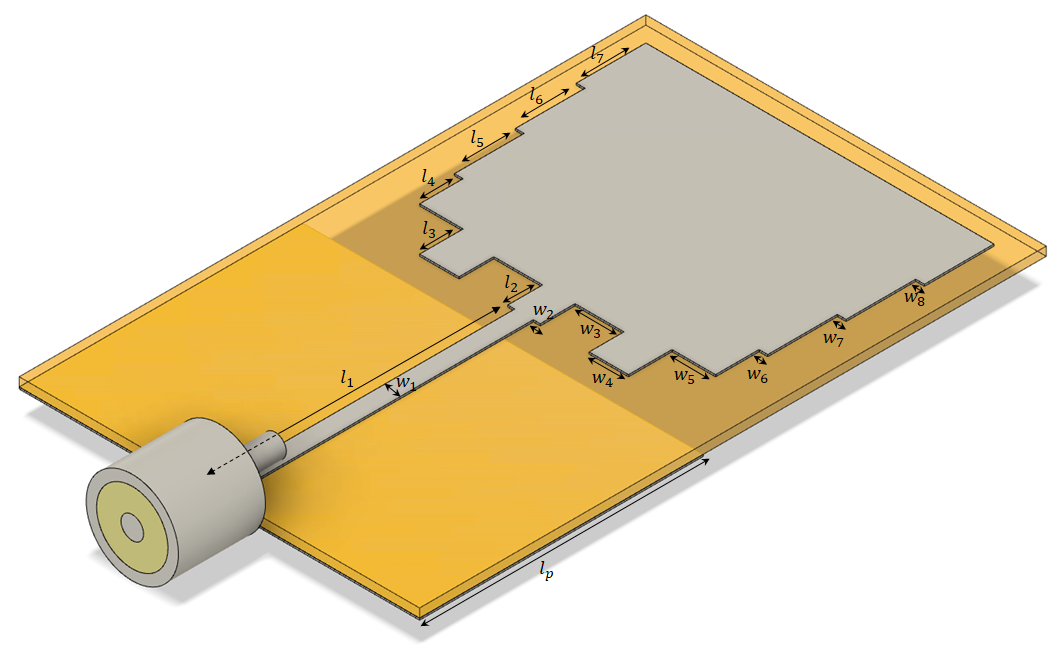}
\caption{Configuration of the single proposed microstrip antenna.}
   \label{Single_ant}
 \end{figure}

 \begin{figure}[!hbt]
\centering
\includegraphics[width=0.5\textwidth]{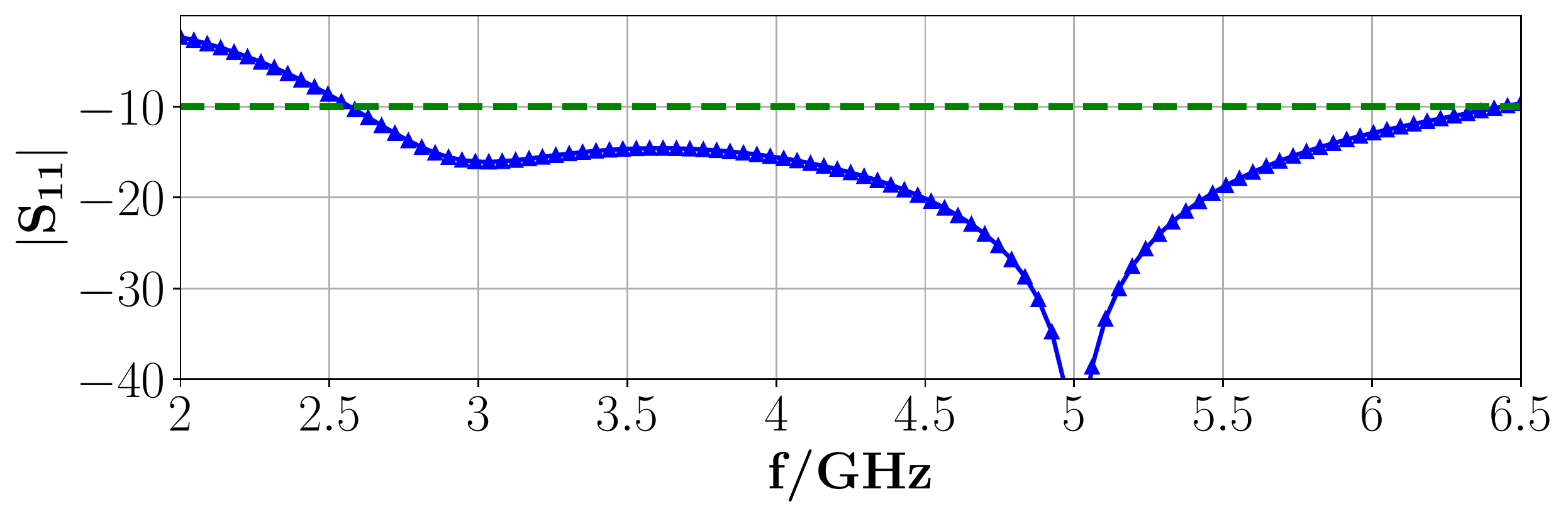}
\caption{Simulated S-parameters of the original antenna. }
   \label{S11_single}
 \end{figure}
 \section{Antenna Design}
\begin{figure}[!t]
\centering
\includegraphics[width=0.275\textwidth]{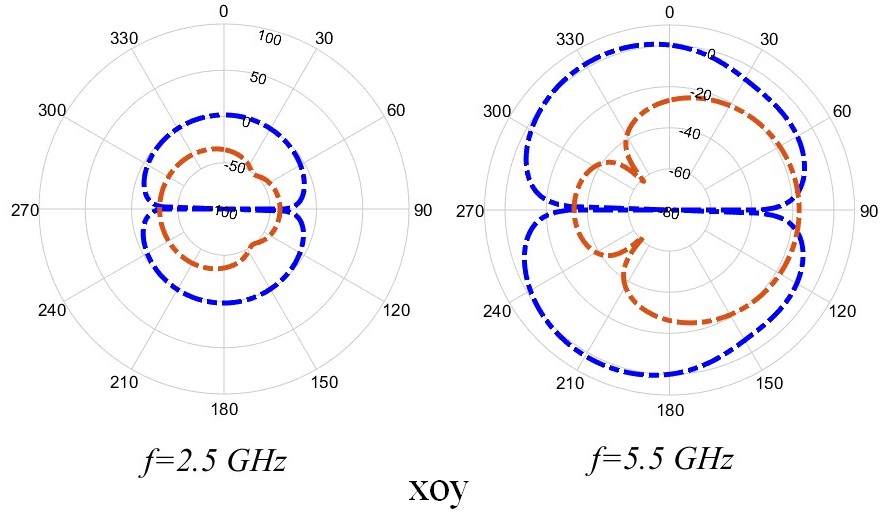}
\includegraphics[width=0.275\textwidth]{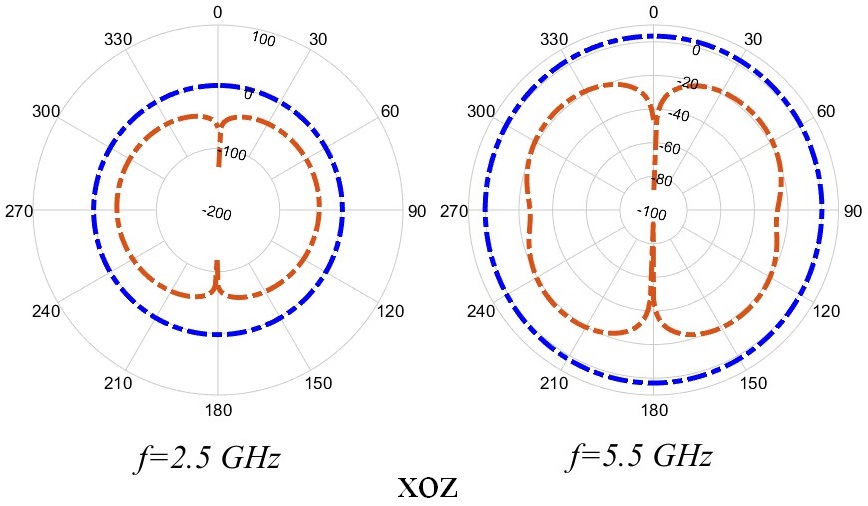}
\includegraphics[width=0.275\textwidth]{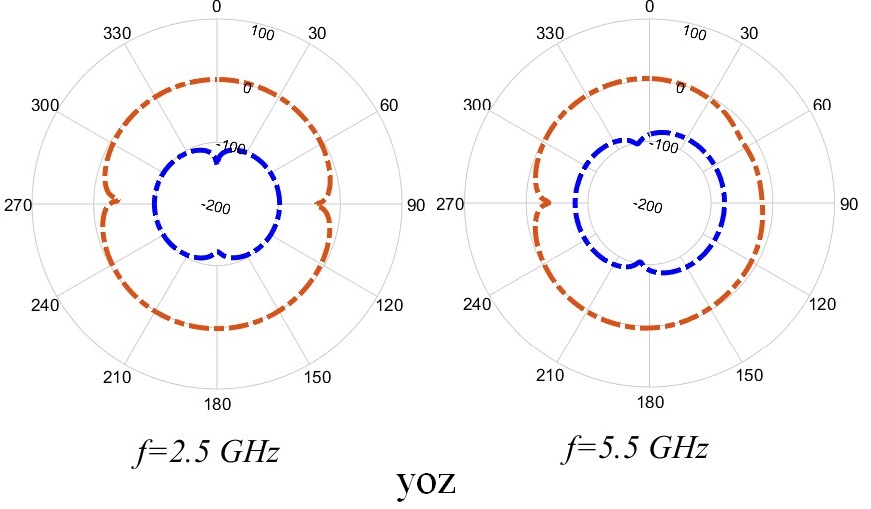}
\caption{Simulated radiation patterns of the single proposed antenna. Blue dash lines represent Phi gain, and red dash lines represent theta gain.}
   \label{singlepattern}
 \end{figure}

The single antenna's configuration used to construct the two ports MIMO antenna is shown in Fig. \ref{Single_ant}. The design is imprinted on a Rogers 4350 substrate with a dielectric constant of 3.66, a loss tangent of 0.004, and a thickness of \SI{0.508}{mm}, with a width and length of \SI{36}{mm} ($W$)$\times$\SI{2}{mm} ($L$). The antenna is coaxially fed, and the optimized antenna dimensions are listed in Table I. A stepped impedance resonator (SIR) structure is employed here to design the radiating element. It implies cascading high-impedance and low-impedance transmission lines from the feed line to the radiate microstrip \cite{SIRM}. Therefore, a wider bandwidth can be achieved.
\begin{table}[h!]
\centering
\caption{Dimensions of the original antenna units: $mm$}
\begin{tabular}{c c c c c c} 
 \hline\hline
 Variable & Value & Variable & Value & Variable & Value \\ [1ex] 
 \hline
 $l_{1}$ & 17.5 & $l_{7}$ & 4     &  $w_{5}$ & 2.5\\ 
 $l_{2}$ & 2    & $l_{p}$ & 16.3  &  $w_{6}$ & 0.5 \\
 $l_{3}$ & 2.5  & $w_{1}$ & 1.07  &  $w_{7}$ & 0.5 \\
 $l_{4}$ & 2.5  & $w_{2}$ & 0.4   &  $w_{8}$ & 0.5 \\
 $l_{5}$ & 4    & $w_{3}$ & 2.8   &          &  \\
 $l_{6}$ & 4    & $w_{4}$ & 2.265 &          & \\   [1ex] 
 \hline\hline
\end{tabular}
\label{table:1}
\end{table}
The preliminary design of the shown microstrip antenna has been evaluated and optimized by using the commercial software CST Studio Suite 2022. The simulated return loss in dB of the initial design of the single antenna is shown in Fig. \ref{S11_single}. It can be seen that the return loss is lower than \SI{-10}{dB} from $2.5 \sim 6.5$ GHz, as expected by employing the SIR structure that delivers an ultra-wideband performance. 
The radiation patterns of the simulated antenna are given in Fig. \ref{singlepattern} in vertical and paralleled planes corresponding to the frequencies of \SI{2.5}{GHz}, \SI{5.5}{GHz}, respectively.
\section{Design of the Dual-Band MIMO Antenna}
\subsection{Without the EBG Loading}
\begin{figure}[!t]
\centering
\includegraphics[width=0.55\textwidth]{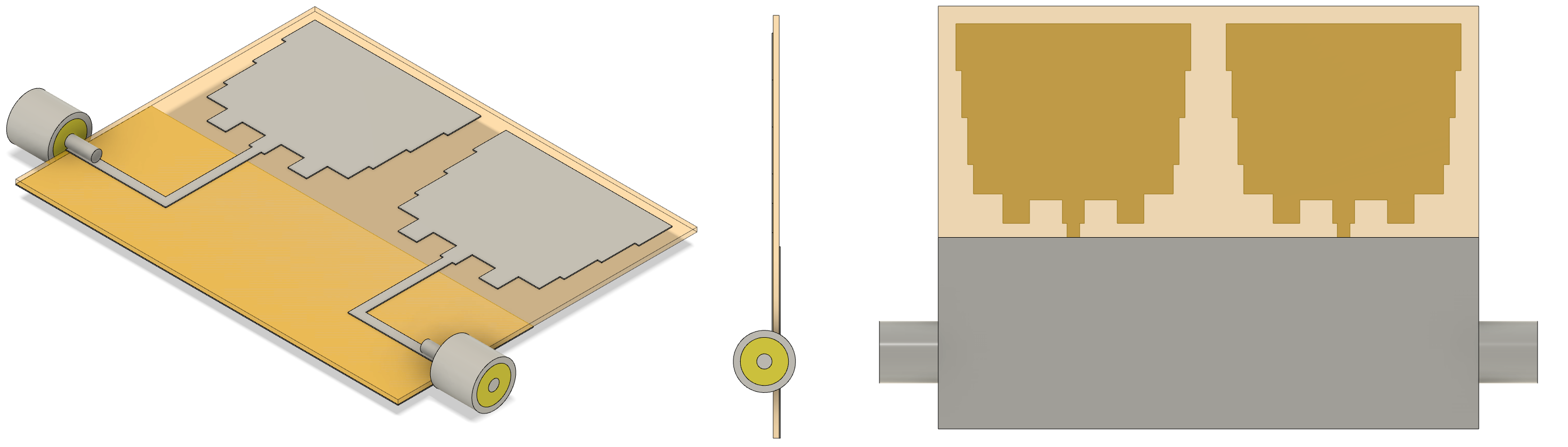}
\caption{Configuration of the MIMO antenna. From left to right: general view, side view, and bottom view. }
   \label{strMIMO}
 \end{figure}

\begin{figure}[!t]
\centering
\includegraphics[width=0.5\textwidth]{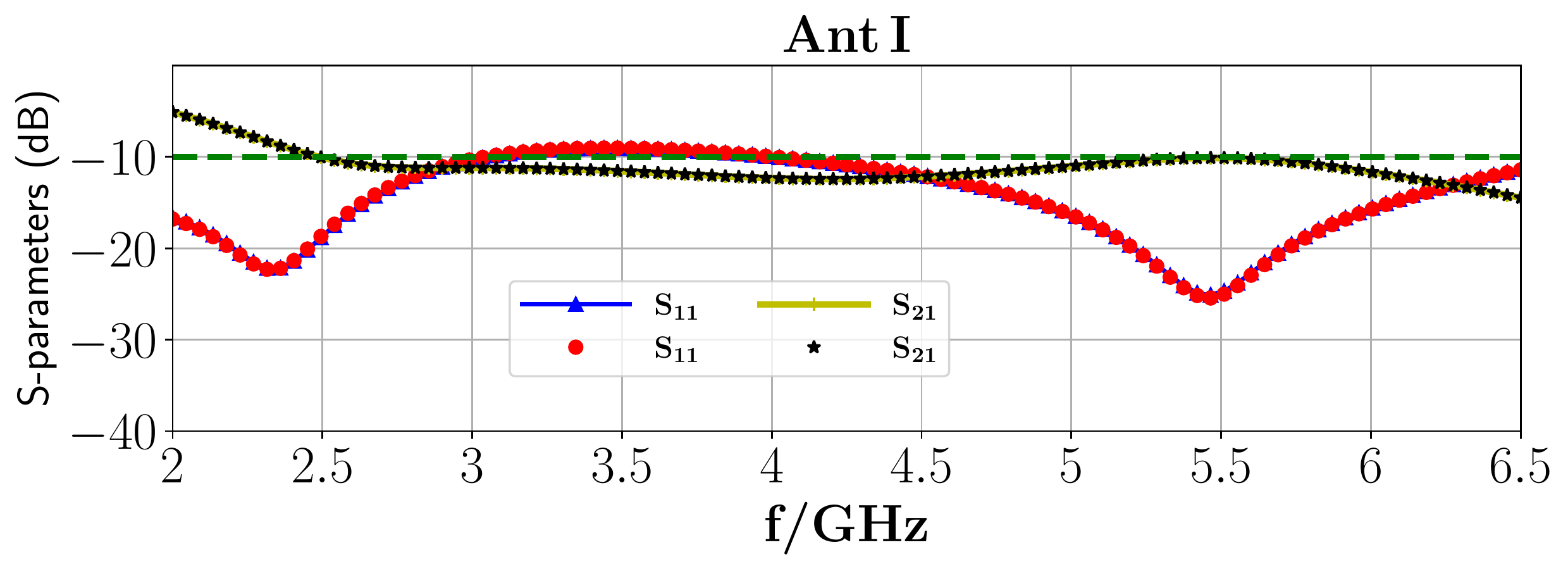}
\caption{Simulated S-parameters of the Ant I.}
\label{SWEBG}
\end{figure}

The general schematic of the ultra-wideband MIMO antenna (Ant I) is shown in Fig. \ref{strMIMO}, in addition to the side view and back view. The distance between the two radiating elements is $0.25\lambda$ at the \SI{2.5}{GHz} and $0.58\lambda$ at \SI{5.8}{GHz}. A microwave bend is used to place the coaxial feed at the side of the MIMO antenna. It is observed the small capacitance that is usually a result of bending the transmission line just slightly changes the antenna performance. The S-parameters of the MIMO antenna is represented in Fig. \ref{SWEBG}. The return loss is under \SI{-10}{dB} in $2.0\sim 3.07$ GHz, and $4.18 \sim 6.5$ GHz. The isolation between two radiating elements is about \SI{10}{dB} in all frequency bands.
 \begin{figure}[!t]
\centering
\includegraphics[width=0.5\textwidth]{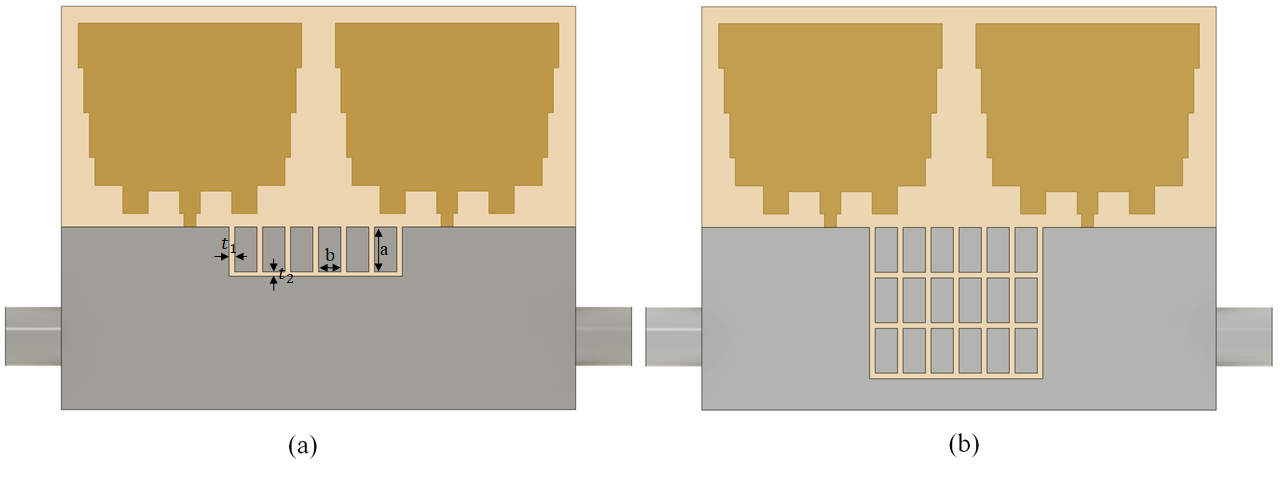}
\caption{Backside of the MIMO antenna with EBG. (a) one row, and (b) three rows of EBG loading. }
   \label{strMIMOEBG}
 \end{figure}
\subsection{With the EBG Loading}
\begin{figure}[!t]
\centering
\includegraphics[width=0.5\textwidth]{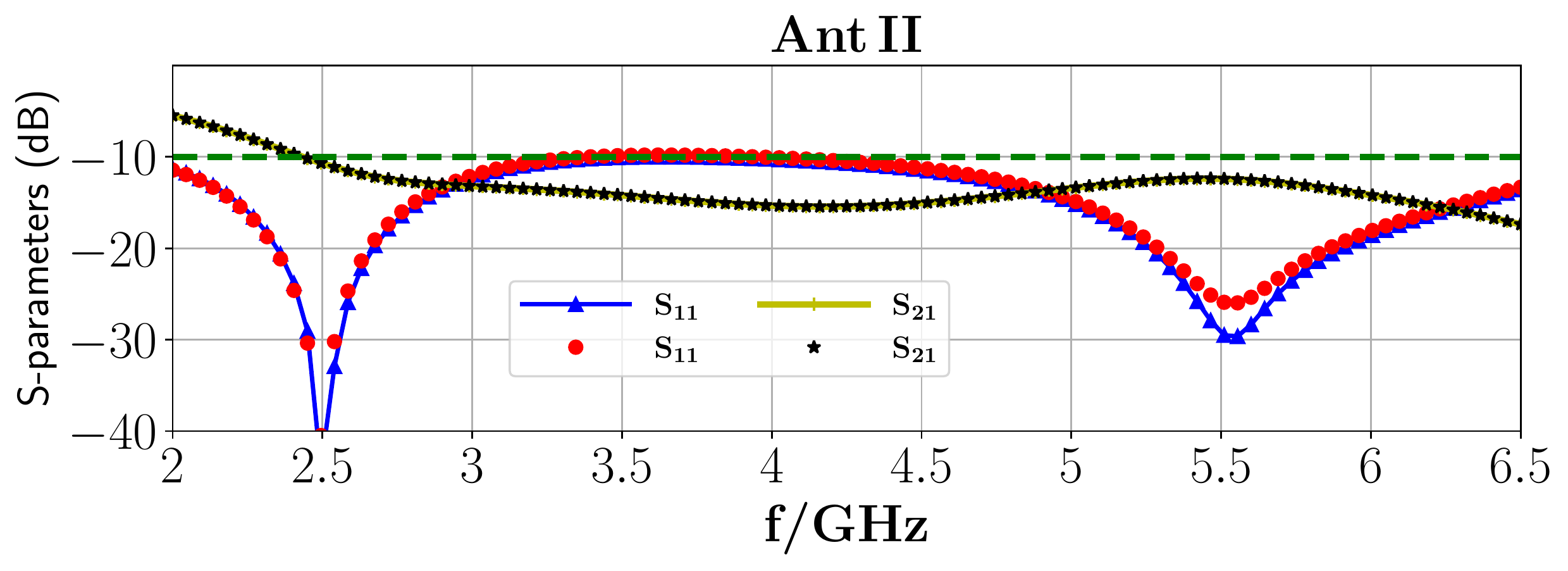}
\caption{Simulated S-parameters of the Ant II. }
\label{SEBG}
\end{figure}

\begin{figure}[!t]
\centering
\includegraphics[width=0.5\textwidth]{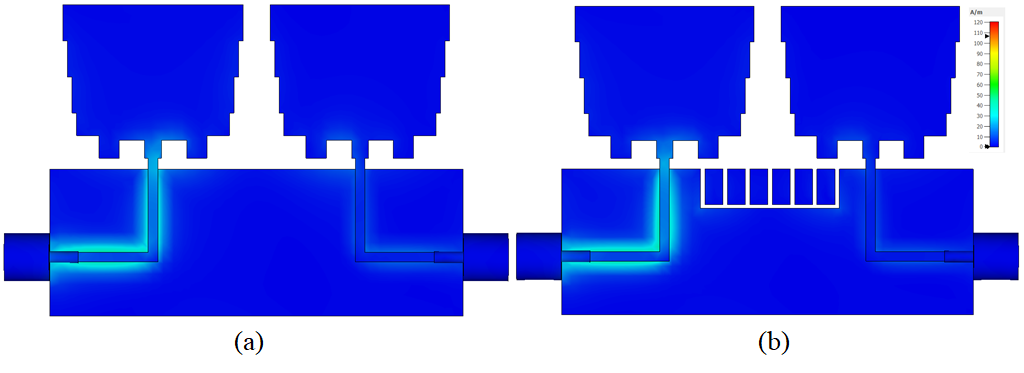}
\caption{Simulated current distribution of the MIMO antenna at 2.5 GHz: (a) without the EBG and (b) with one later of the EBG. }
   \label{Surface1}
 \end{figure}

\begin{figure}[!t]
\centering
\includegraphics[width=0.5\textwidth]{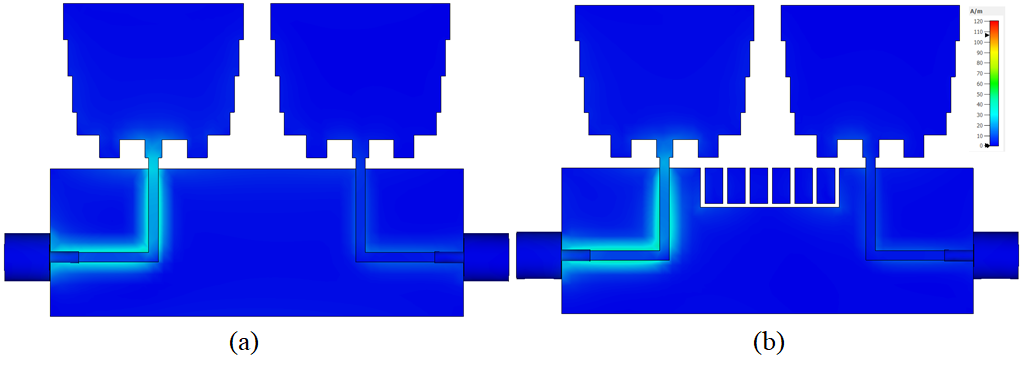}
\caption{Simulated current distribution of the MIMO antenna at 5.5 GHz: (a) without the EBG and (b) with one later of the EBG. }
   \label{Surface2}
 \end{figure}

\begin{figure}[!t]
\centering
\includegraphics[width=0.5\textwidth]{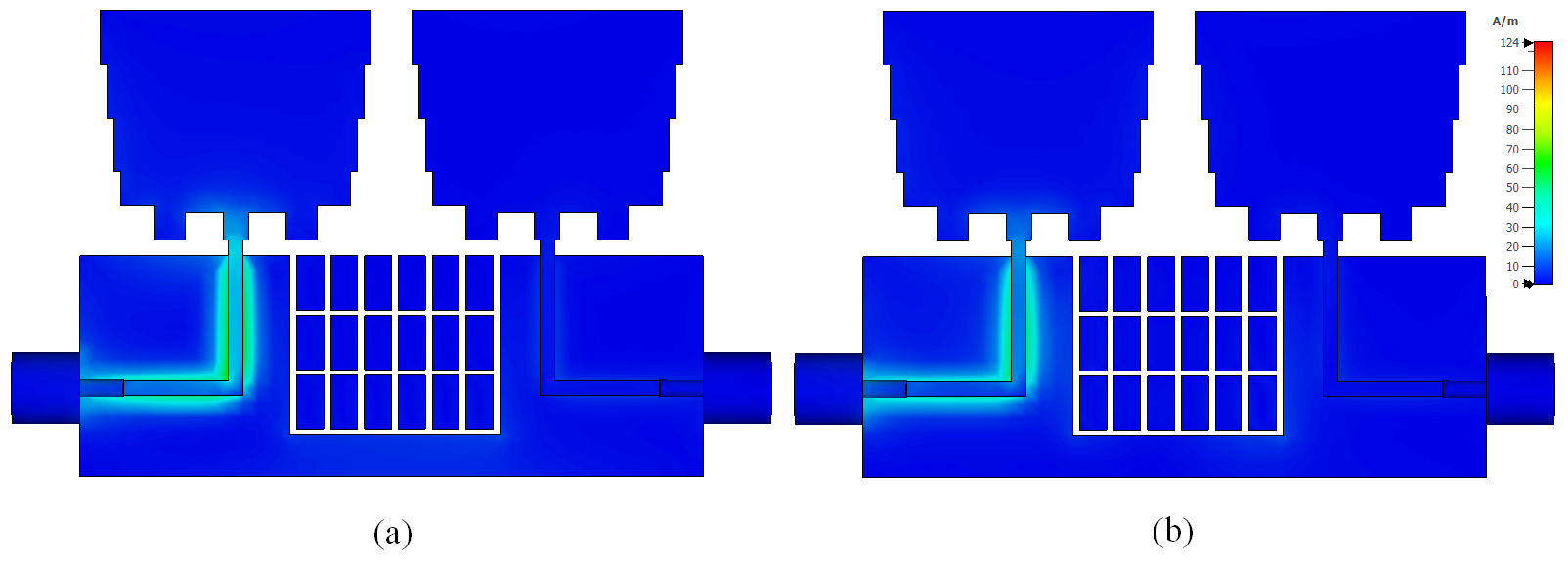}
\caption{Simulated current distribution of the MIMO antenna with three later of the EBG at (a) \SI{2.5}{GHz} and (b) \SI{5.5}{GHz}. }
   \label{Surface3}
 \end{figure}

\begin{figure}[!t]
\centering
\includegraphics[width=0.475\textwidth]{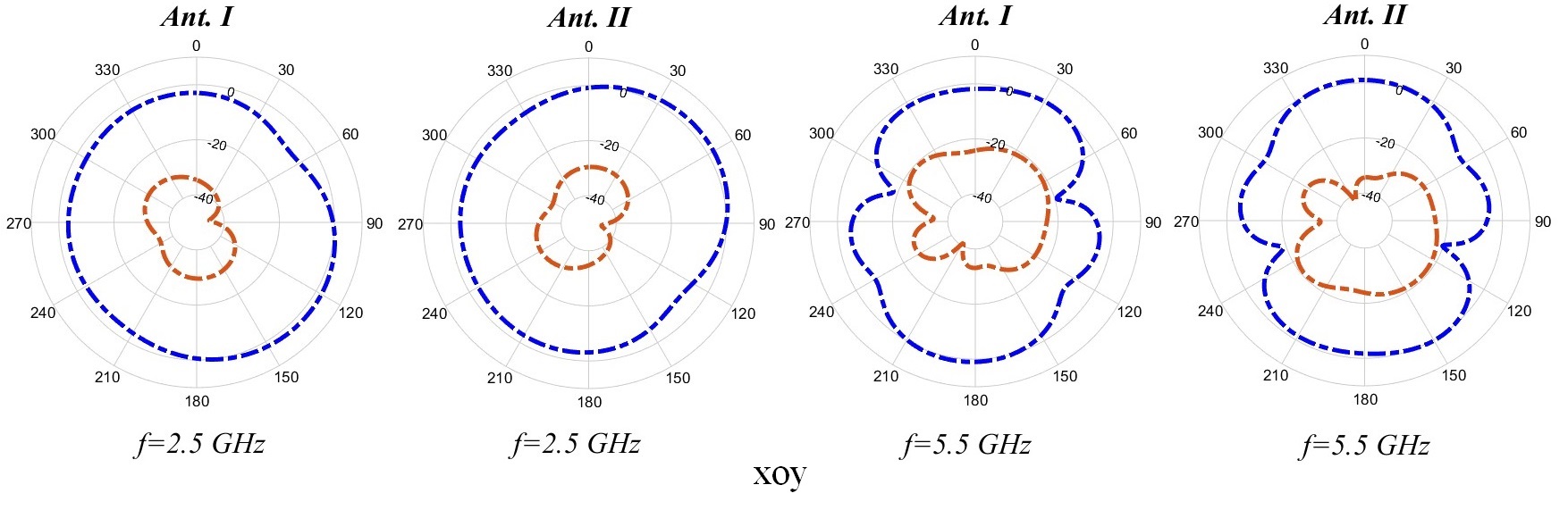}
\includegraphics[width=0.475\textwidth]{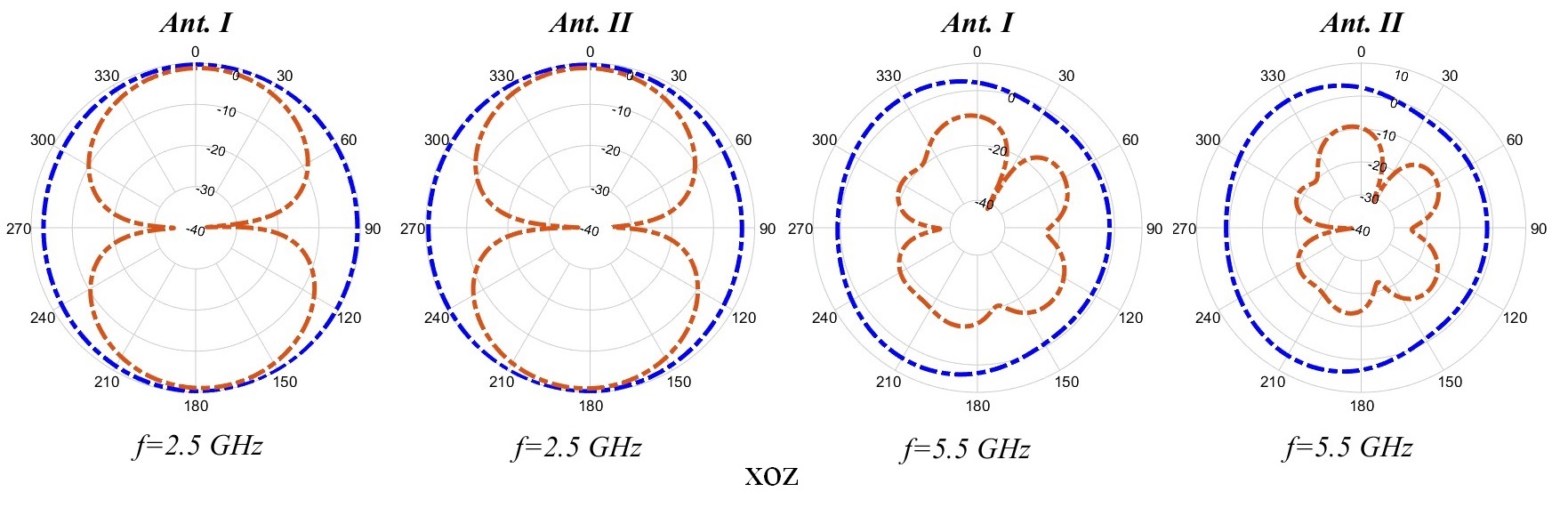}
\includegraphics[width=0.475\textwidth]{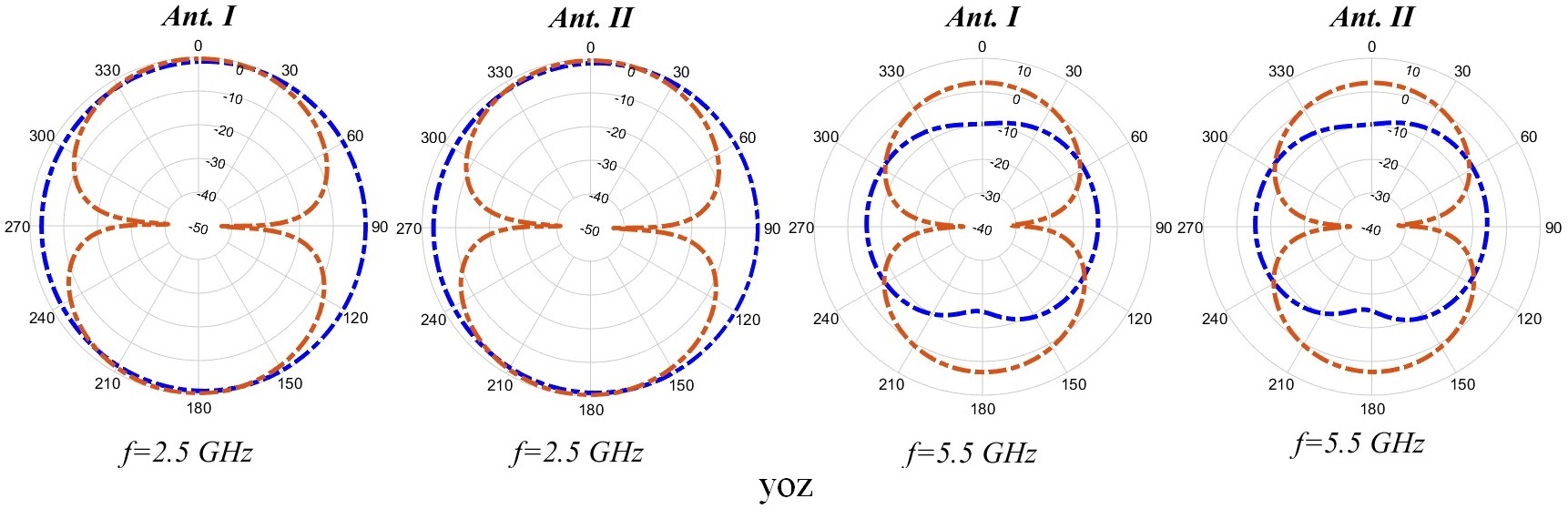}
\caption{Simulated radiation patterns of the proposed decoupled
MIMO antenna (Ant II). Blue dash lines represent Phi gain, and red dash lines represent theta gain. }
   \label{MIMOpattern}
 \end{figure}

\begin{figure}[!t]
\centering
\includegraphics[width=0.5\textwidth]{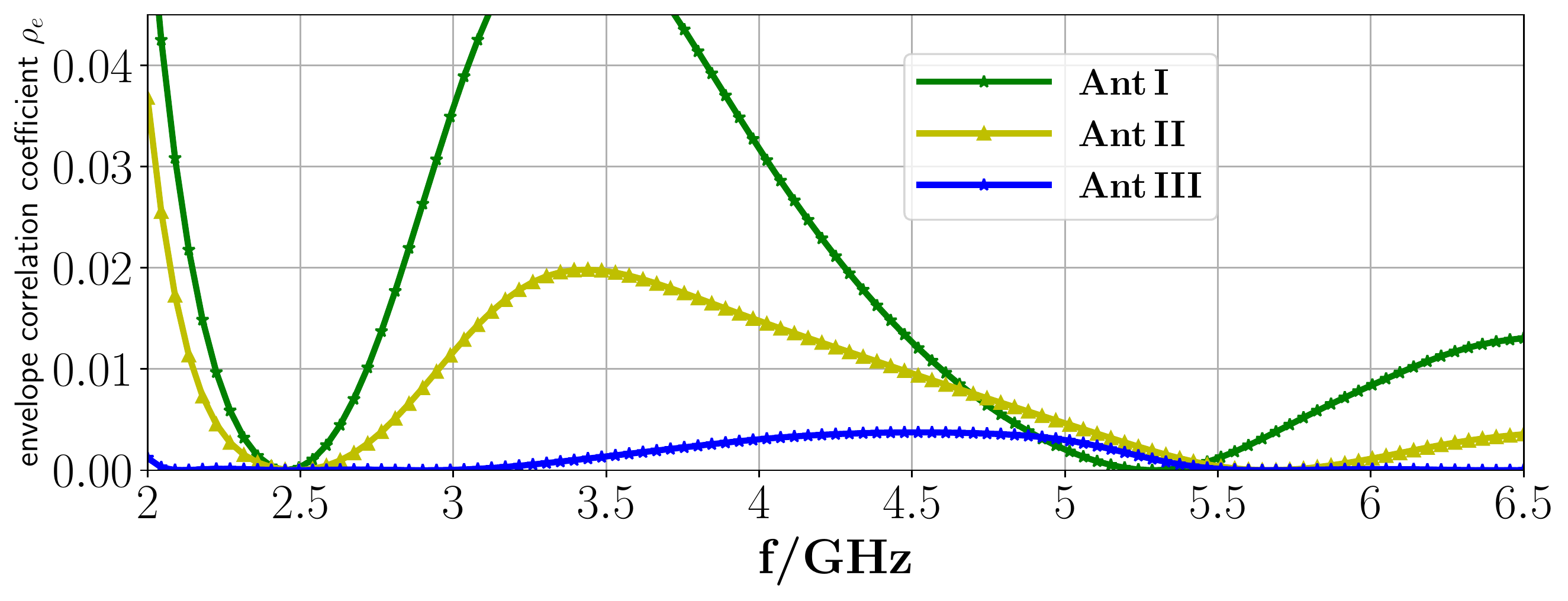}
\caption{Simlulated ECC ($\rho_{e}$) calculated from S-parameters for different MIMO antennas. }
   \label{ECCf}
 \end{figure}

By adding EBG to the MIMO antenna, the structure shown in Fig. \ref{strMIMOEBG} is obtained. Fig. \ref{strMIMOEBG} (a) represents one row of the EBG loading (Ant II), and Fig. \ref{strMIMOEBG} represents three rows of the EBG loading (Ant II). The unit cell of the planar EBG employed here is semi-periodic metallic patches printed on the same side of the ground, however, it doesn't cover all the width of the ground. The optimized EBG unit cell's dimensions
are listed in Table II. 
\begin{table}[h!]
\centering
\caption{Dimensions of the EBG's unit cell: $mm$}
\begin{tabular}{c| c c c c} 
 \hline\hline
 Variable & $a$  & $b$ & $t_{1}$ &  $t_{2}$ \\ 
 Value    & $4$  & $2$ & $0.5$   &  $0.4$  \\
   [1ex] 
 \hline\hline
\end{tabular}
\label{table:1}
\end{table}
The $S_{11}$ and $S_{21}$ of the first and second radiating elements of the Ant II with one row of EBG loading are illustrated in Fig. \ref{SEBG}. The S-parameters of the MIMO antennas represent satisfactory behavior in the entire frequency band of $2.0 \sim 6.5$ GHz, compared to the Ant I without the EBG loading, which can cover WiMax ($2.5/3.5/5.5$ GHz), WLAN ($5.2/5.8$-GHz), Sub-6G, as shown in Fig. \ref{SEBG}. The isolation between two ports of the MIMO antenna is about \SI{13}{dB} in almost all frequency bands which shows \SI{3}{dB} improvement compared to the Ant I. \newline
\indent In order to find the influences on the resonant frequencies
and bandwidths of corresponding structural parameters of the EBG, a parametric study has been fulfilled. By increasing the width of the single cell of the EBG, and keeping constant the height of that, the resonance frequency of the lower band ($2.4 \sim 3.0$ GHz) almost remains constant. However, the resonance frequency of the higher band ($5.0 \sim 6.5$ GHz) decreases and gets wider. The isolation  ($S_{21}$) between two ports also increases by increasing the width of the unit cell. By increasing the length of the unit cell in a fixed width, the resonance frequency in the lower band increases, and the resonance frequency of the higher band decreases and gets wider.\newline
\begin{figure}[!t]
\centering
\includegraphics[width=0.5\textwidth]{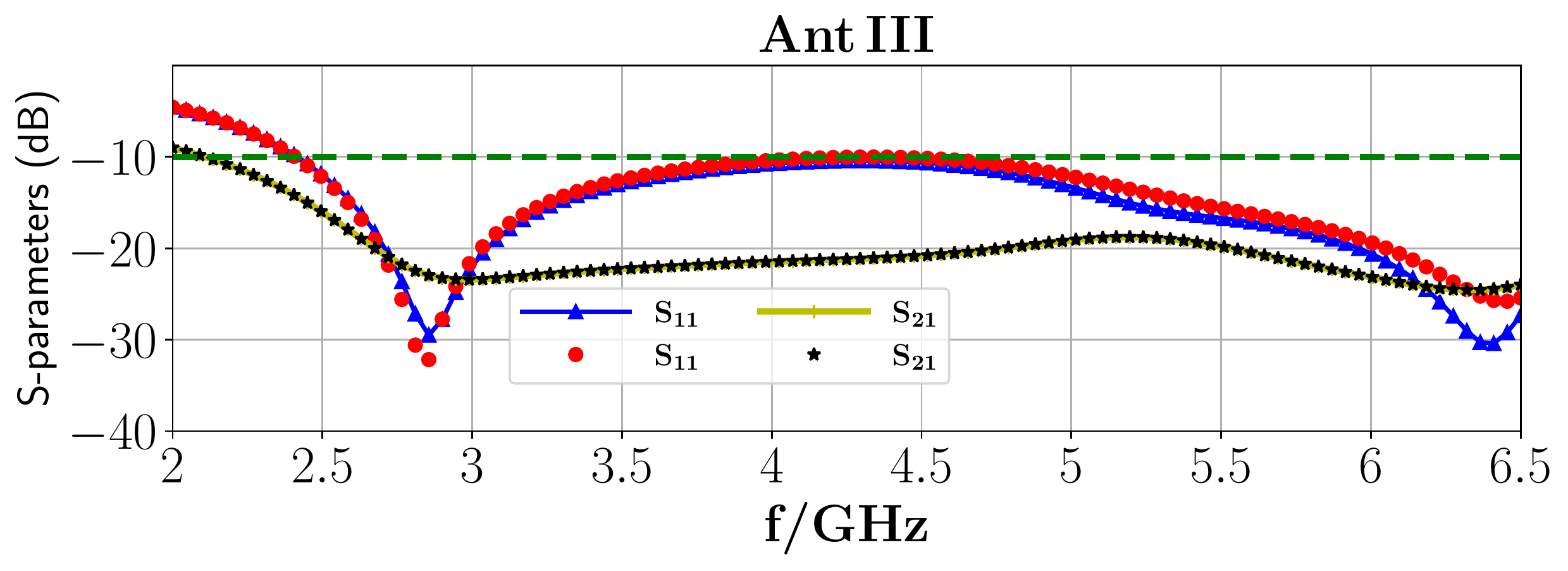}
\caption{Simulated S-parameters of the Ant III.}
\label{SEBG2}
\end{figure}
The isolation between two ports of the previously introduced MIMO antenna can be further improved. In this respect, two more rows of semi-periodic EBG loading are added to the ground of the antenna (Ant III), as shown in Fig. \ref{strMIMOEBG} (b). Integrating more rows of the EBG loading (in both $x$ and $y$ direction) into the MIMO antenna will curb the current distribution between two transmission lines and result in improving the isolation. The S-parameters of the first and second radiating elements are shown in Fig. \ref{SEBG2}. Compared to the S-parameters of the Ant II with one row of EBG, \SI{6}{dB} improvement at \SI{2.5}{GHz}, and \SI{8}{dB} improvement at \SI{5.5}{GHz} can be obtained. However, the return loss in structure (b) is under \SI{-10}{dB} from \SI{2.4}{GHz} rather than \SI{2}{GHz} in structure (a), which still is covering the mentioned applications' operating frequencies.\newline
\indent The surface current of the proposed MIMO antenna with/without EBG at the frequencies of \SI{2.5}{GHz} and \SI{5.5}{GHz} are shown in Fig. \ref{Surface1}, and \ref{Surface2}, \ref{Surface3}, to better compare the effect of the EBG. The figures depict that the coupling at the \SI{2.5}{GHz} and \SI{5.5}{GHz} bands have been interrupted by using the EBG. Moreover, an improvement can be observed in the radiating. Increasing the EBG loading in $y$-direction has led to improve the isolation as shown in Fig. \ref{Surface3}. \newline
To better represent the working mechanism of the proposed MIMO antenna, the radiation patterns of the simulated Ant II are given in Fig. \ref{MIMOpattern} in vertical and paralleled planes corresponding to the operating frequencies of \SI{2.5}{GHz}, and \SI{5.5}{GHz}, respectively. It can be seen that the patterns are quasi-omnidirectional, which is suitable for wireless communication terminals to receive signals from any direction.\newline
\indent In the MIMO communication system, the ECC parameter represents how much the channel is uncorrelated to obtain if the designed antenna is appropriate for the MIMO application. Here, we use the S-parameter method to calculate the ECC of the designed MIMO antenna. Less correlation (more channel capacity) between the MIMO antenna's elements can be found by lowering the value of the ECC. The ECC is calculated as follows
\begin{equation}
    \rho_e=\frac{\Big|S^{*}_{11}S_{21} + S^{*}_{22}S_{12}\Big|^{2}}{\Big|(1-|S_{11}|^{2}-|S_{21}|^{2})(1-|S_{22}|^{2}-|S_{12}|^{2})\Big|}
\end{equation}
The ECC of the proposed MIMO antennas with and without the EBG is illustrated in Fig. \ref{ECCf}. As can be seen from \ref{ECCf}, the ECC of  Ant II and Ant III is lower than 0.02 at $2.0 \sim 6.5$ GHz, which means a good MIMO performance. However, the ECC of Ant III with more EBG loading is clearly better than Ant II without one row of EBG.

 \section{Conclusion}
In this manuscript, an ultra-wideband microstrip antenna with integrated EBG loading operating at $2.0 \sim 6.5$ GHz WLAN and Sub-6G bands is designed and analyzed. Two microstrip antennas, designed using the SIR technique, were unified to construct two ports MIMO antenna. The isolation of the proposed MIMO antenna has been improved to \SI{20}{dB} by employing EBG loading at the ground plane. The MIMO antenna represents a quasi-omnidirectional pattern, which is suitable for wireless communication. The proposed MIMO antenna has a compact size, lower envelope correlation coefficient, and low cost and can be used in different wireless devices.
\section*{Acknowledgment}

\IEEEtriggercmd{}
\IEEEtriggeratref{4}

\bibliography{references}

\begin{thebibliography}{10}

\bibitem{spacediv}
M.~P. Karaboikis, V.~C. Papamichael, G.~F. Tsachtsiris, C.~F. Soras, and V.~T.
  Makios, ``Integrating compact printed antennas onto small diversity/mimo
  terminals,'' {\em IEEE Transactions on Antennas and Propagation}, vol.~56,
  no.~7, pp.~2067--2078, 2008.

\bibitem{Patterndiv}
E.~Rajo-Iglesias, O.~Quevedo-Teruel, M.~L. Pablo-Gonzalez, and M.~P.
  Sanchez-Fernandez, ``Performance of mimo systems employing multiple compact
  patch antennas with radiation pattern diversity,'' in {\em 2006 First
  European Conference on Antennas and Propagation}, pp.~1--6, 2006.

\bibitem{polardiv}
L.~Garcia-Garcia, B.~Lindmark, and C.~Orlenius, ``Design and evaluation of a
  compact antenna array for mimo applications,'' in {\em 2006 IEEE Antennas and
  Propagation Society International Symposium}, pp.~313--316, 2006.

\bibitem{EBG1}
F.~Yang and Y.~Rahmat-Samii, ``Microstrip antennas integrated with
  electromagnetic band-gap (ebg) structures: a low mutual coupling design for
  array applications,'' {\em IEEE Transactions on Antennas and Propagation},
  vol.~51, no.~10, pp.~2936--2946, 2003.

\bibitem{Fconverted}
J.~Deng, J.~Li, L.~Zhao, and L.~Guo, ``A dual-band inverted-f mimo antenna with
  enhanced isolation for wlan applications,'' {\em IEEE Antennas and Wireless
  Propagation Letters}, vol.~16, pp.~2270--2273, 2017.

\bibitem{Lshape}
Z.~Li, Z.~Du, M.~Takahashi, K.~Saito, and K.~Ito, ``Reducing mutual coupling of
  mimo antennas with parasitic elements for mobile terminals,'' {\em IEEE
  Transactions on Antennas and Propagation}, vol.~60, no.~2, pp.~473--481,
  2012.

\bibitem{Neutral}
H.-L. Peng, R.~Tao, W.-Y. Yin, and J.-F. Mao, ``A novel compact dual-band
  antenna array with high isolations realized using the neutralization
  technique,'' {\em IEEE Transactions on Antennas and Propagation}, vol.~61,
  no.~4, pp.~1956--1962, 2013.

\bibitem{Tstub}
J.~Sui and K.-L. Wu, ``A general t-stub circuit for decoupling of two dual-band
  antennas,'' {\em IEEE Transactions on Microwave Theory and Techniques},
  vol.~65, no.~6, pp.~2111--2121, 2017.

\bibitem{Lumped}
M.~M. Albannay, J.~C. Coetzee, X.~Tang, and K.~Mouthaan, ``Dual-frequency
  decoupling for two distinct antennas,'' {\em IEEE Antennas and Wireless
  Propagation Letters}, vol.~11, pp.~1315--1318, 2012.

\bibitem{DGS1}
S.~Soltani, P.~Lotfi, and R.~D. Murch, ``A dual-band multiport mimo slot
  antenna for wlan applications,'' {\em IEEE Antennas and Wireless Propagation
  Letters}, vol.~16, pp.~529--532, 2017.

\bibitem{DGS2}
M.~S. Sharawi, A.~B. Numan, M.~U. Khan, and D.~N. Aloi, ``A dual-element
  dual-band mimo antenna system with enhanced isolation for mobile terminals,''
  {\em IEEE Antennas and Wireless Propagation Letters}, vol.~11,
  pp.~1006--1009, 2012.

\bibitem{SIRM}
Z.~H. Ma and Y.~F. Jiang, ``L-shaped slot-loaded stepped-impedance microstrip
  structure uwb antenna,'' {\em Micromachines}, vol.~11, no.~9, 2020.

\end{thebibliography}

\end{document}